\documentclass[aps]{revtex4}
\usepackage{amssymb}
\usepackage{amsmath}
\usepackage{graphicx}
\usepackage{epsfig,amsmath}

\begin{document}

\title{Dynamics of coreless vortices and rotation-induced dissipation peak
in superfluid films on rotating porous substrates}
\author{S. K. Nemirovskii$^{1}$ and E. B. Sonin$^{2}$}
\affiliation{$^{1}$Institute of Thermophysics, Lavrentyev ave., 1, 630090, Novosibirsk,
Russia\\
$^{2}$Racah Institute of Physics, Hebrew University of Jerusalem, Givat Ram,
Jerusalem 91904, Israel}
\date{\today }

\begin{abstract}
We analyze dynamics of 3D coreless vortices in superfluid films covering
porous substrates. The 3D vortex dynamics is derived from the 2D dynamics of
the film. The motion of a 3D vortex is a sequence of jumps between
neighboring substrate cells, which can be described, nevertheless, in terms
of quasi-continuous motion with average vortex velocity. The vortex velocity
is derived from the dissociation rate of vortex-antivortex pairs in a 2D
film, which was developed in the past on the basis of the
Kosterlitz-Thouless theory. The theory explains the rotation-induced
dissipation peak in torsion-oscillator experiments on $^4$He films on
rotating porous substrates and can be used in the analysis of other
phenomena related to vortex motion in films on porous substrates.

PACS-numbers: 67.40.Vs, 67.57.Fg, 67.70.+n, 67.40.Rp
\end{abstract}

\maketitle

\section{Introduction}

Superfluid $^{4}$He films adsorbed in porous media is an actual topic in
physics of superfluidity\cite{Rp}. Studying of this system gives a unique
possibility to investigate the interplay between 2D and 3D physics,
especially the character of the transition to the superfluid state. On one
side, torsion-oscillator experiments reveal the dissipation peak near the
temperature of the superfluid onset $T_{c}$, which is predicted by the
dynamical theory of vortex-antivortex pairs \cite{AHNS} based on the
Kosterlitz-Thouless theory for 2D films \cite{KT}. On another side, they
found that in films on porous substrates the superfluid density critical
index $\sim 2/3$ \cite{Rp} of the 3D system and the sharp cusp of the
specific heat \cite{Murphy} at $T_{c}$ are similar to those near the $%
\lambda $ transition of the bulk $^{4}$He.

An important insight into physics of superfluid films in porous media is
provided by torsion-oscillator experiments with rotating substrates. The
porous substrate is usually modeled with a \textquotedblleft jungle
gym\textquotedblright\ structure \cite{Mahta,Min,GW,OK}: a 3D cubic lattice
of intersecting cylinders of diameter $a$ with period $l$ (Fig.\ref%
{fig1}). Multiple connectivity of superfluid films in porous media
allows a variety of vortex configurations, and probably most important from
them is a coreless, or pore 3D vortex, which is just a flow around the
vortex pores having nonzero circulation. Due to the presence of a new type
of topological defects, one could expect an essential difference in the
response between the plane film and the porous-medium film under rotation.
This expectation was confirmed by torsional-oscillator experiments, which
revealed a rotation-induced peak in dissipation (inverse quality factor) as
a function of temperature. The additional peak was shifted from the
stationary (static) peak that was observed without rotation \cite%
{Fukuda_JLTP_98,Fukuda04}. Double-peak structure essentially differs from
the case of the plane film, where the only effect of rotation was to broaden
the stationary peak \cite{AdGl,Rpflt}.

A semi-empirical interpretation of the rotation-induced dissipation peak was
suggested in Ref. \onlinecite{Fukuda04}. It is clear that the rotation can
affect dissipation via rotation contribution to the velocity field. So it is
a nonlinear correction to the response. Instead of the derivation of such a
correction from the theory, the authors of Ref. \onlinecite{Fukuda04} used
the data on the nonlinear response taken from the independent experiment on
large-amplitude torsion-oscillations. This provided a qualitative
explanation of the rotation peak, and even of some quantitative features of
it, but could not pretend to be a full theory of the effect. The present
work suggests really a theory of the rotation peak deriving the parameters
of the peak from the parameters of the film and the substrate. The key role
in our scenario is played by the 3D coreless vortices. We derived dynamics
of these vortices from dynamics of 2D films adsorbed in porous media. It was
suggested \cite{Fukuda04} that motion of the coreless 3D vortices occurs in
a creeping manner by jumping from cell to cell. These jumps are related to
dissociation of the vortex-antivortex pair on one side of a rod separating
different pores, with subsequent annihilation of the pair on another side of
a rod. The result of the jump is a shift of velocity circulation to a
neighboring pore. Though our analysis was focused on application to the
torsion-oscillator experiments, it is valid for description 3D vortex motion
in many other cases, at least in those, where the vortex moves to distances
much larger than the average period of the substrate structure. An important
example is steady vortex motion in zero-frequency experiments.

In Sec. \ref{2-3D} we describe how dynamics of 3D coreless vortices is
connected with dynamics of 2D films covering the substrate. This provides a
bridge between the effective-continuous-medium 3D description and dynamics
of 2D films. Section \ref{PD} reviews the theory of dissociation of
vortex-antivortex pairs on the basis of the Kosterlitz-Thouless theory. In
Sec. \ref{CP} the pair-dissociation rate is analyzed near the critical
temperature, where there is an analytical solution of Kosterlitz's recursion
equations. The theory is applied to the analysis of the torsion-oscillation
experiments in Sec.~\ref{TO}. The rotation-induced dissipation peak directly
follows from the theory, and its position, shape and even height are in a
good agreement with the experiment. Section \ref{concl} is devoted to
conclusions.

\section{ Coreless vortices: from 2D to 3D vortex dynamics}

\label{2-3D}

\begin{figure}[tbp]
\includegraphics[width=8.6cm]{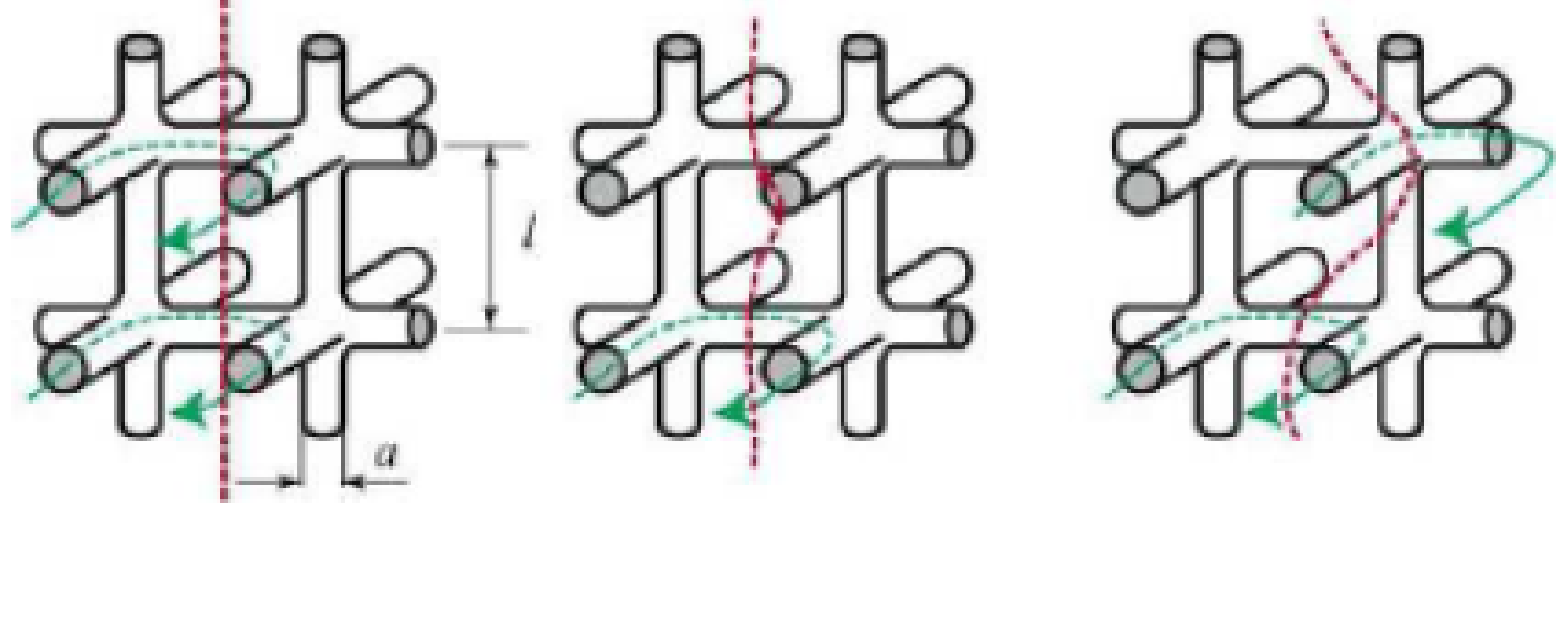}
\caption{(Color online) Vortex creep in the jungle gym structure: the coreless vortex
(dashed line) crosses a cylinder between two cells via dissociation of the
vortex-antivortex pair. The arrowed curved lines show circulation around the
cells where the vortex line is located.}
\label{fig1}
\end{figure}

\begin{figure}[tbp]
\includegraphics[width=15cm]{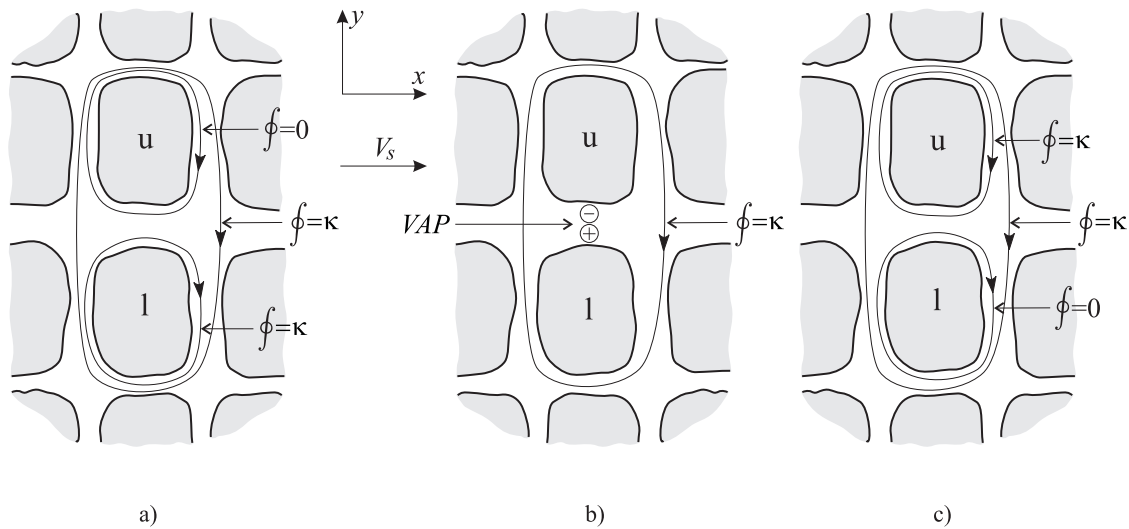}
\caption{Vortex creep. The superfluid moves with velocity $%
\mathbf{V}_s$ along the axis $x$, and the vortex creeps along the axis $y$.
(a) The state before the jump. There is circulation $\protect\kappa $ around
the lower cell (l) as it is shown in picture. The same circulation exists
around any path inclusive cell (l). The circulation around two cells (l) and
(u) is shown. The coreless vortex is located at the cell (l). (b) The vortex
on the way between cells (l) and (u). There is the vortex-antivortex pair
(VAP) in the film. Circulations around the paths inclusive only of cell (l)
or (u) are not defined, while for the path around the two cells circulation
is equal to $\protect\kappa $ as before. (c) The state after the jump. The
coreless vortex is located at the cell (u). No circulation around the cell
(l) anymore, but there is circulation around the cell (u), or around the two
of them.}
\label{fig2}
\end{figure}

In a continuous medium a vortex is a topological defect with nonzero
circulation around the vortex axis. There is an area around the vortex line
with a suppressed order parameter, which is called vortex core. But topology
of a porous medium allows circulation of superfluid velocity around a pore
\emph{without suppression of the order parameter anywhere inside the
superfluid film}. This leads to the concept of a coreless vortex. The vortex
``line'' in this case is not a line at all; this is a chain of the jungle
gym structure cells with nonzero circulation around them. Schematically the
coreless vortex is depicted in Fig.~\ref{fig1}.
This coreless structure of a vortex rules out usual type of vortex motion in
a continuous medium simply because the coreless vortex has no continuous
coordinate: its position is discrete and is determined by a cell with
nonzero circulation around it. The only way for the vortex to move is to
jump from cell to cell. They call such a type of vortex motion \emph{vortex
creep}. But any jump is in fact a process in time (whatever short) during
which the vortex line inevitably cross a rod covered by a superfluid film.
So during the jump the ``coreless'' vortex does have cores: at the place
where it enters the rod and at the place where it comes out from the rod
(Fig.~\ref{fig1}). These two cores together form a 2D
vortex-antivortex pair, which should grow, dissociate, and eventually
annihilate on the other side of the rod. This process lead to a discrete
shift of the vortex line to a neighboring cell. For better illustration a
two-dimensional picture of this process is shown in Fig. \ref{fig2}. It
shows two cells: upper (u) and lower (l). Before the process (Fig. \ref%
{fig2}a) there is circulation $\kappa =h/m_{He}$ around the lower cell as
shown in the figure. Provided that there is no other coreless vortices
nearby, the same circulation exists around any path inclusive of the cell
(l) as shown for the path around the both two cells. The transient process
of the ``jump'' from the cell (l) to the cell (u) is shown in Fig. \ref%
{fig2}b. The vortex-antivortex pair is present in the film between two
pores. This makes circulation around any of the two cells undefined: it
depends on whether the path goes outside or inside the pair. Only
circulation around the two cells together remains equal to $\kappa$. Figure %
\ref{fig2}c shows the state after the process: the coreless vortex is now
located in cell (u). Though this scenario is shown in the plane picture, it
is directly applicable to the 3D jungle gym structure with the 2D vortex and
antivortex moving around a cylindric rod.
In reality creation and dissociation of vortex-antivortex pairs is a
stochastic process, which is determined by the average dissociation rate $%
R(T,v_{s})$ (number of dissociation events per second and per unit area of a
film), which depends on temperature $T$ and, most important for us, on
average superfluid velocity $v_{s}$ in the film. In our case the superfluid
velocity field consists of two parts: $\mathbf{v}_{s}=\mathbf{v}_{c}+\mathbf{%
V}_{s}$, where $\mathbf{v}_{c}$ originates from a circular flow around the
vortex line at rest and $\mathbf{V}_{s}$ is a transport superfluid velocity
with respect to a moving substrate. In Fig. \ref{fig2} the velocity $%
\mathbf{V}_{s}$ is directed along the axis $x$. So it it is added to the
circular velocity $\mathbf{v}_{c}$ above the cell with circulation $\kappa $
and is subtracted from $\mathbf{v}_{c}$ below this cell. Therefore the jumps
of the vortex line up and down are unbalanced and result in some average
drift (creep) of the vortex line up the picture in Fig. \ref{fig1}. This
jump-like process can, nevertheless, be described with an average vortex
velocity determined as
\begin{equation}
V_{L}=Al(R(v_{c}+V_{s},T)-R(v_{c}-V_{s},T))\approx 2Al\frac{\partial R}{%
\partial v_{s}}V_{s}~,  \label{VLy}
\end{equation}%
where $l$ is the period of the jungle gym structure, and $A$ is the area of
the rod separating neighboring pores. According to this scenario of vortex
motion the vortex velocity $\mathbf{V}_{L}$ is strictly normal to the
velocity $\mathbf{V}_{s}$.

Let us compare the relation (\ref{VLy}) with the general relation connecting
the vortex velocity of the vortex line with the normal and the superfluid
velocities \cite{Sonin_RMP}:
\begin{equation}
\mathbf{V}_{L}=\mathbf{V}_{s}+\alpha\hat{z}\times ({\mathbf{V}}_{n}-{\mathbf{%
V}}_{s})-\alpha ^{\prime }({\mathbf{V}}_{n}-{\mathbf{V}}_{s})\,.
\label{VL_standard}
\end{equation}%
In the problem under consideration we can assume that the normal component
is clamped to the porous-glass substrate oscillating with the velocity $%
\mathbf{V}_{g}$. So their velocities coincide: $\mathbf{V}_{n}=\mathbf{V}%
_{g} $. In fact equation (\ref{VLy}) is written for the system moving with
the substrate, where $\mathbf{V}_{n}=\mathbf{V}_{g}=0$. Eventually Eqs. (\ref%
{VLy}) and (\ref{VL_standard}) agree if
\begin{equation}
\alpha =2Al\frac{\partial R}{\partial v_{s}},\ \ \ \ \ \ \ \alpha ^{\prime
}=-1\,.\ \ \ \   \label{a,a'}
\end{equation}%
Note that the condition $1+\alpha ^{\prime }=0$ providing that vortices move
normally to the superfluid motion means the absence of the \textquotedblleft
effective\textquotedblright\ Magnus force \cite{Magn}, which is defined as
the term $\propto \lbrack \hat{z}\times \mathbf{V}_{L}]$ is the balance of
forces on the vortex. In superconductors this leads to the total absence of
the Hall effect. Absence of the effective Magnus force is typical for
lattice systems \cite{magn}, in contrast to uniform continuous media with
Galilean invariance. The crossover between these two cases was recently
studied by numerically solving the Gross-Pitaevskii equation \cite{GK}.

So we have demonstrated that the creep of coreless vortices, which is
realized via sequences of discrete jumps from cell to cell, can be described
in the terms usually used for 3D vortices moving in a continuous medium.
Still the parameters of this 3D \textquotedblleft
effective-medium\textquotedblright\ description must be determined within
the theory of 2D films covering the multi-connected substrate. The crucial
parameter to be determined is the derivative $dR/dv_{s}$ of the dissociation
rate $R$.\newline

\section{ Rate of pair dissociation}

\label{PD}

The thermally activated dissociation of vortex-antivortex pair was analyzed
by Ambegaokar \emph{et al.} \cite{AHNS} (see also references to later works
in Ref. \onlinecite{GilBow}) on the basis of the Kosterlitz-Thouless theory.
They considered superfluid films on plane substrates, while in our case
films cover cylindrical surfaces. But as we shall see below, the relevant
scale (size of the pair at the saddle point) is small compared to the
substrate curvature and the curvature may be ignored.

The pair dissociation is accompanied by overcoming the potential barrier.
The barrier corresponds to the saddle point of the vortex-pair energy as a
function of the radius-vector $\mathbf{r}$ connecting vortex with
antivortex,
\begin{equation}
U({\mathbf{r}},{\mathbf{v}}_{s})=\frac{\rho _{s0}\kappa ^{2}}{2\pi }%
\,\int_{a_{0}}^{r}\frac{d^{2}\mathbf{r}}{\tilde{\epsilon}(r)r}-\rho
_{s0}\kappa \mathbf{r}\cdot (\mathbf{v}_{s}\times \hat {z})\,.
\label{potential}
\end{equation}%
Here $\rho_{s0}$ is the bare superfluid density, $\tilde \epsilon$ is
Kosterlitz's static scale-dependent dielectric constant determined from the
integral equation
\begin{equation}
{\frac{1}{\tilde \epsilon(r)}}-1=-{\frac{\pi \rho_{s0}\kappa^2 }{T}}
y_0^2\int_{r_0}^r dr{\frac{r^3}{r_0^4}}\exp{\left[-{\frac{ \rho_{s0}\kappa^2
}{2\pi T}}\int_{r_0}^r {\frac{dr }{\epsilon(r)r}}\right]} \, ,
\label{epsilon}
\end{equation}
$r_{0}$ is the core radius of the 2D vortex, $y_0=e^{-E_0/T}$, and $E_0$ is
the energy of the vortex core. The dielectric constant $\tilde \epsilon(r)$
takes into account screening of interaction between a vortex and an
antivortex at distance $r$ by pairs of smaller size. In the limit $r\to
\infty$ the dielectric constant determines the ratio of the bare and the
renormalized superfluid densities: $\rho_s=\rho_{s0}/\tilde \epsilon(\infty)$%
.

The integral equation (\ref{epsilon}) can be reduced to Kosterlitz's
recursion equations:
\begin{equation}
\frac{dK(l)}{dl}=-4\pi ^{3}y^{2}K^{2},\ \ \ \frac{dy^2(l)}{dl}=(4-2\pi K)
y^2\, .  \label{KT}
\end{equation}
Here $l=\ln (r/r_{0})$, the dielectric constant is replaced by the ratio $%
\tilde \epsilon(l)=K(0)/K(l)$, where $K(0)=\rho _{s0}\kappa ^{2}/4\pi
^{2}k_{B}T$ is the bare Kosterlitz -Thouless coupling constant related to
the bare superfluid density $\rho _{s0}$, and
\begin{equation}
y(l)= y_0\exp{\left[2l-\pi \int_0 ^{l }K(l^{\prime}) dl^{\prime}\right]}
\end{equation}
is the rescaled activity.

Solving the Fokker-Planck equation for the distribution of vortex-antivortex
pairs, Ambegaokar \emph{et al.} \cite{AHNS} obtained the following
expression for the dissociation rate [see Eq. (4.6) in their article]
\begin{equation}
R=\frac{2D}{r_{s}^{4}}y^{2}(l_{s})\exp (2\pi K(l_{s})).  \label{rate_AHNS}
\end{equation}%
Here $D$ is the vortex diffusion coefficient, $r_{s}$ is the saddle-point
value of the pair size $r$, and $l_{s}=r_{s}/r_{0}$ is its logarithm. The
saddle point for the effective potential (\ref{potential}) is reached when $%
\mathbf{r}$ is perpendicular to $\mathbf{v}_{s}$ and $r_{s}$ satisfies the
condition
\begin{equation}
\frac{\kappa }{2\pi \tilde{\epsilon}(r_{s})r_{s}}=\frac{\kappa K(0)}{2\pi
K(l_{s})r_{s}}=v_{s}.  \label{saddle point}
\end{equation}%
If the velocity $v_{s}$ is small the values of $l_{s}$ and $r_{s}$ are
large, and this condition yields that
\begin{equation}
r_{s}=\frac{\kappa K(0)}{2\pi K(\infty )v_{s}}\,.  \label{saddle pointL}
\end{equation}

In order to find the mutual friction parameter $\alpha $ from Eq. (\ref{a,a'}%
), we need the derivative of the dissociation rate with respect to the
velocity $v_{s}$. Using Eqs. (\ref{KT}) and (\ref{saddle point}) we obtain:
\begin{equation}
{\frac{dR}{dv_{s}}}={\frac{8\pi ^{2}DK(0)}{\kappa r_{s}^{3}}}%
y^{2}(l_{s})\exp [2\pi K(l_{s})]\,.  \label{R'}
\end{equation}

In summary, Eqs (\ref{VLy}), (\ref{a,a'}) and (\ref{R'}) show how the creep
motion of the coreless vortices in a porous medium is described in terms of
parameters determining motion of 3D quantum vortices in continuous media.
This description can be used for various problems related to vortex motion
in porous media.

\section{Pair dissociation near the critical point}

\label{CP}

For understanding the nature of the rotation dissipation peak we need to
study the temperature dependence of the dissociation rate at temperatures
close to the critical one. At these temperatures Kosterlitz's recursion
equations have an analytical solution \cite{AHNS}. One can introduce a small
$x(l)=\pi \lbrack K(l)-K_{c}(\infty )]$, where $K_{c}(l)$ yields values of $%
K(l)$ at the critical point and at $l\rightarrow \infty $ $K_{c}(\infty
)=2/\pi $. Then the recursion relations can be written as
\begin{equation}
\frac{dx(l)}{dl}=-(4\pi y)^{2},\ \ \ \frac{dy^{2}(l)}{dl}=-2xy^{2}\,.
\label{KTlin}
\end{equation}%
Their solution is
\begin{equation}
x(l)=x_{\infty }\coth \left( x_{\infty }l+\coth ^{-1}{\frac{x_{0}}{x_{\infty
}}}\right) =x_{\infty }\frac{x_{0}\cosh (x_{\infty }l)+x_{\infty }\sinh
(x_{\infty }l)}{x_{0}\sinh (x_{\infty }l)+x_{\infty }\cosh (x_{\infty }l)}\,,
\label{K(Tl)}
\end{equation}%
\begin{equation}
4\pi y(l)=x_{\infty }\mbox{csch}\left( x_{\infty }l+\coth ^{-1}{\frac{x_{0}}{%
x_{\infty }}}\right) =\frac{4\pi y_{0}x_{\infty }}{x_{0}\sinh (x_{\infty
}l)+x_{\infty }\cosh (x_{\infty }l)}\,.  \label{y(Tl)}
\end{equation}%
Here $x_{0}=x(0)$ and $x_{\infty }=x(\infty )$. The solution satisfies the
condition
\begin{equation}
x_{\infty }^{2}=x(l)^{2}-[4\pi y(l)]^{2}=x_{0}^{2}-(4\pi y_{0})^{2}\,.
\label{CR}
\end{equation}%
At the critical point $x_{\infty }=0$ and Eqs. (\ref{K(Tl)}) and (\ref{y(Tl)}%
) become
\begin{equation*}
x_{c}(l)\approx {\frac{x_{0}}{1+x_{0}l}}~,~~4\pi y_{c}(l)\approx {\frac{x_{0}%
}{1+x_{0}l}}~.
\end{equation*}%
According to Eq. (\ref{CR}), at the critical point the parameters $x_{0}$
and $y_{0}$ satisfy the relation
\begin{equation}
x_{0c}=\pi \left[ K_{c}(0)-{\frac{2}{\pi }}\right] =(4\pi y_{0c})^{2}\,.
\end{equation}%
As usually assumed in the 2D vortex dynamics, the bare superfluid density
does not vary near the critical point. Then, since $K(0)\propto 1/T$ and $%
y_{0}=e^{-E_{0}/T}$, their dependence on the relative temperature $%
t=(T_{c}-T)/T_{c}$ is
\begin{equation}
K(0)\approx K_{c}(0)(1+t)~,~~y_{0}\approx y_{0c}\left( 1-{\frac{E_{0}}{T_{c}}%
}t\right)~.
\end{equation}%
Then Eq. (\ref{CR}) yields that at $t>0$ $x_{\infty }=2b\sqrt{t}$ with $%
b=16\pi e^{-E_{0}/T_{c}}[1+2\pi (1+E_{0}/T_{c})e^{-E_{0}/T_{c}}]$. This
leads to the square-root cusp
\begin{equation}
\rho _{s}(T)=\rho _{s}(T_{c})(1+b\sqrt{t})  \label{rho_s(t)}
\end{equation}
in the critical behavior of the renormalized superfluid density, which was
revealed by Nelson and Kosterlitz \cite{NK} with numerical calculations.

Using all these relations together with the assumption that $v_{s}$ so low
that at the saddle point $l_{s}\gg x_{0c}$, we obtain linear dependence of
the dissociation rate on $t$:
\begin{equation}
{\frac{dR}{dv_{s}}}={\frac{DK_{c}(0)}{2\kappa r_{s}^{3}l_{s}^{2}}}%
e^{4}(1-\gamma t)\,,  \label{R'_1}
\end{equation}

where
\begin{equation*}
\gamma =2b^{2}l_{s}+{\frac{4b^{2}}{3}}-1\,.
\end{equation*}

\section{Torsional oscillations of rotating porous substrate and comparison
with the observed rotation dissipation peak}

\label{TO}

Let us now apply the theory to oscillatory motion of the substrate
superimposed on its steady rotation with the angular velocity $\mathbf{%
\Omega }$. So the substrate velocity field is $[\mathbf{\Omega }\times
\mathbf{r}]+\mathbf{V}_{g}$. Only the oscillatory component $\mathbf{V}%
_{g}\propto e^{-i\omega t}$ is important for us. Using the Euler equation
\begin{equation}
{\frac{\partial \mathbf{V}_{s}}{\partial t}}+[2\mathbf{\Omega }\times
\mathbf{V}_{L}]=-\mathbf{\nabla \mu }
\end{equation}%
and Eq. (\ref{VL_standard}) with $\alpha ^{\prime }=-1$ one obtains for an
oscillatory components of the velocities (the chemical potential $\mu$ is
not relevant for the azimuthal motion):
\begin{equation}
\mathbf{V}_{s}={\frac{2\Omega \alpha }{i\omega +2\Omega \alpha }}\mathbf{V}%
_{g}\,.
\end{equation}%
This relation determines the drag of the superfluid component by the
oscillating substrate. Analyzing now the balance of forces for the torsional
resonator, as was done many times in the past, one obtains the following
contribution to the inverse quality factor of the torsion oscillator for the
slow rotation $\Omega \alpha \ll \omega $:
\begin{equation}
\Delta Q^{-1~}=\frac{V\rho _{s3}}{M}{\frac{2\Omega \alpha }{\omega }}\,,
\end{equation}%
where $M$ is the total mass of the torsional oscillator, $V$ is the total 3D
volume (including pores), and $\rho _{s3}$ is the effective 3D superfluid
mass density in the porous-glass substrate of the volume $V$, which is
connected with the 2D superfluid density $\rho _{s}$ of the film by the
relation
\begin{equation}
\rho _{s3}={\frac{\rho _{s}}{V}}~.  \label{rho_3D}
\end{equation}%
Here $A_{tot}$ is the total area of the film. For the \textquotedblleft
jungle gym\textquotedblright\ structure with cell size $l$ and the rod
diameter $a$
\begin{equation}
{\frac{A_{tot}}{V}}\approx {\frac{\pi a}{l^{2}}}~.
\end{equation}%
On the other hand, in the theory of 2D superfluid films starting from Ref. %
\onlinecite{AHNS} they describe the drag of the superfluid component
introducing the \textquotedblleft dynamical dielectric
constant\textquotedblright\ $\epsilon (\omega )$ (one should not mix it up
with Kosterlitz's static dielectric constant $\tilde\epsilon$ introduced in
Sec. \ref{PD}!) :
\begin{equation}
\mathbf{V}_{s}=\left[ 1-{\frac{1}{\epsilon (\omega )}}\right] \mathbf{V}%
_{g}\,.
\end{equation}%
Then
\begin{equation}
\Delta Q^{-1~}=-\frac{V\rho _{s3}}{M}\mbox{Im}{\frac{1}{\epsilon }}=-\frac{%
A_{tot}\rho _{s}}{M}\mbox{Im}{\frac{1}{\epsilon }}  \label{Q_key}
\end{equation}%
with the imaginary part of the inverse dielectric constant equal to
\begin{equation}
\mbox{Im}{\frac{1}{\epsilon }}=-{\frac{2\Omega \alpha }{\omega }}=-{\frac{%
4Al\Omega }{\omega }}\ {\frac{\partial R}{\partial v_{s}}}\,.
\label{inverse epsilon}
\end{equation}

Substituting Eqs. (\ref{rho_s(t)}) and (\ref{R'_1}) into Eqs. (\ref{Q_key})
and (\ref{inverse epsilon}), we see that the square-root cusp in the
superfluid density is crucial for its temperature dependence in the critical
area and for existence of the rotation dissipation peak:
\begin{equation}
\Delta Q^{-1}={\frac{4Al\Omega }{\omega }}\frac{A_{tot}}{M}\rho _{s}{\frac{%
\partial R}{\partial v_{s}}}=\Delta Q^{-1}(T_{c})(1+b\sqrt{t}-\gamma t)\,.
\label{peak}
\end{equation}%
The factor $\gamma \propto l_{c}$ is expected to be large, but the the
square-root cusp is more essential at small $t$, and the inverse quality
factor has a maximum at $t=b^{2}/4\gamma ^{2}$ rather close to the critical
point.

The linear approximation used for derivation of Eqs. (\ref{K(Tl)}) and (\ref%
{y(Tl)}) is more or less truthful only for $\left\vert T-T_{c}\right\vert $
not exceeding $0.005$ K. Though the dissipation maximum at $T\approx 0.6234$
K is in this interval (see below), the low temperature side of the
dissipation peak is not, and the peak width cannot be determined using the
analytical formulas (\ref{K(Tl)}) and (\ref{y(Tl)}). Thus though the
analytical expression (\ref{peak}) qualitatively explains the observed
dissipation peak itself, it is not sufficient for its quantitative
description. A more accurate quantitative analysis required numerical
calculations.

Let us now gather from Refs. \cite{Fukuda_JLTP_98} and \cite{Fukuda04} all
quantitative data needed for comparison with the theory. The porous-glass
substrate can be modeled with the jungle-gym structure with the diameter of
rods $a\approx 1~\mu $m and the structure period $l\approx 2.5~\mu $m. Then
the circulation velocity around the pore is estimated as $v_{s}=\kappa
/4l\approx 1$ cm/sec. According to Ref. \cite{Fukuda04}, the transition
temperature is $T_{c}$ $\approx 0.628$ K. The areal density of superfluid
component at the critical temperature (jump of density) can be evaluated
from the Kosterlitz-Thouless relation
\begin{equation*}
\rho _{s}=\frac{8\pi k_{B}T_{c}}{\kappa ^{2}}=\allowbreak 2.\,\allowbreak
179\,1\times 10^{-9}\ \mbox{g/cm}^{2}.
\end{equation*}%
We also need the bare areal superfluid density $\rho _{s0}$. It can be
calculated from dependence of the transition temperature on the thickness:
According to Ref. \onlinecite{Fukuda_JLTP_98}, the coverage of the substrate
is about $33\times 10^{-10}\ $ mol/cm$^{2}$. Then $\rho _{s0}\approx
1.32\times 10^{-8}$ g/cm$^{2}$, and the bare Kosterlitz -Thouless coupling
constant $K(0)$ is
\begin{equation*}
K(0)=\frac{\rho _{s0}\kappa ^{2}}{4\pi ^{2}k_{B}T}\approx 3.8~.
\end{equation*}%
Quantity $b$ entering relations (\ref{K(Tl)})-(\ref{y(Tl)}) can be obtained
from the width $\sim 0.02$ K of the experimental curve describing static
dissipation peak of paper \cite{Fukuda04}. It yields $b$ about $3.6975$.
Furthermore, the diffusion coefficient is $D=2.8\times 10^{-7} $ cm$^{2}/$s,
and the core radius of vortices on film can be estimated as $a_{0}\approx
25\times 10^{-8}$ cm.

We can now find the pair size $r_{s}$ at the saddle point from Eq. (\ref%
{saddle pointL}): $r_{s}\approx 2.\,\allowbreak 9\,8\times
10^{-5}\allowbreak $ cm. This corresponds to $\ l_{s}=\ln (r_{s}/r_{0})=4.78$%
. On one hand, these values are large enough in order to justify our
assumption of large $l_{s}$ compared to $x_{0c}$. On the other hand, we see
that $r_{s}<<a$, which justifies our consideration of vortex-antivortex
unbinding using the theory for plane films.

We also need the value of $A_{tot}\rho_{s}(T)/M$ in Eq. (\ref{Q_key}). This
is the ratio of the temperature dependent superfluid mass to the mass of the
empty cell. According to Ref. \onlinecite{Fukuda_JLTP_98} it is
approximately equal to $10^{-5}.$ For determination of the temperature
dependent superfluid mass $\rho _{s}(T)$ we have built the extrapolating
function using the experimental results obtained from the measuring of the
shift of the oscillations period (see Fig. 2 from Ref. \onlinecite
{Fukuda04}).

In order to calculate $K(l,T)$ and $y(l,T)$ numerically we use procedure
proposed in Ref. \onlinecite{Rpflt} (see Appendix A there). In the
temperature interval from $T_{c}$ to the temperature $T\approx 0.6234$ K
corresponding to the dissipation maximum the analytical expressions based on
the linear approximation are more or less truthful. But in the low
temperature region $T<0.6234$ K we performed the numerical calculation using
the Mathematica program. Following Ref. \onlinecite{Rpflt}, for any
temperature $\left\vert T-T_{c}\right\vert $ and corresponding $x_{0}$ we
choose a value $l_{0}$ so that $x_{\infty }l_{0}=\pi /2.$ Under this choice
both $y(l_{0},T)$ and deviation of $K(l_{0},T)$ from $2/\pi $ are rather
small and the analytical expressions (\ref{K(Tl)}) and (\ref{y(Tl)}) are
still good enough for evaluation of $K(l,T)~$\ and $y(l,T).$ Further we take
$K(l_{0},T)~$\ and $y(l_{0},T)$ as initial conditions for numerical
integration (with respect to variable $l$) of the recursive
Kosterlitz-Thouless relations (\ref{KT}). In this way we are able to restore
$K(l_{s},T)~$\ and $y(l_{s},T)$ at saddle point $l_{s}$ and further to find $%
\partial R/\partial v_{s}$ and eventually $\Delta Q^{-1}$ in the whole
temperature interval $T<T_{c}$.
In Fig.\ref{fig3} we plot the theoretical $\Delta Q^{-1}(T)$ (dashed
line) together with the experimental data of Fukuda \emph{et al.} \cite%
{Fukuda04}. The observed rotation induced dissipation peak (the left peak)
was scaled by the angular velocity $\Omega $ reducing it to the value
measured at $\Omega =6.28$ rad/s. The plots for different values of $\Omega $
collapse on the same curve (see Fig. 2 in Ref. \onlinecite{Fukuda04}), which
proves linear dependence of rotation induced dissipation on $\Omega $. For
better comparison of the peak shape we fit the theoretical height of the
peak to the experimental one. But in fact the values of $\Delta Q^{-1}(T)$
at the maxima do not differ essentially: At the angular velocity $\Omega
=6.28$ rad/s they are $3.54\ast 10^{-8}$ in the theory and $2.4\ast 10^{-8}$
in the experiment (i.e., about 70 \% from the theoretical value). Keeping in
mind that nice agreement in the peak shape and position was achieved without
any additional fitting the agreement is really satisfactory.
\begin{figure}[tbp]
\includegraphics[width=10cm]{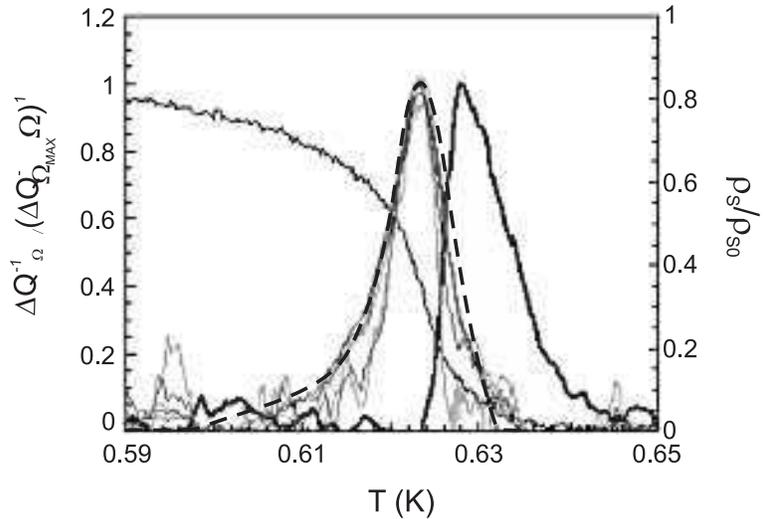}
\caption{Comparison of the experimental and the theoretical inverse quality
factor of torsional oscillations. The right peak is the static peak, which
is present without rotation. The group of the left peaks is rotation-induced
ones obtained experimentally for various $\Omega $ and scaled (divided) by $%
\Omega $. The fact that all these curves collapse well on the same curve
proves linearity on $\Omega $. The dashed line shows the theoretical inverse
quality factor. The vertical scale of it was fitted so that the height of
the theoretical peak coincided with the experimental rotation peak, but the
width and the position of the theoretical peak were calculated without any
fitting. The theoretical curve ends with the critical point. The
experimental superfluid density scaled by its zero temperature value is is
also displayed (the left vertical axis).}
\label{comparison}
\end{figure}
We conclude the quantitative analysis of the torsional-oscillation
experiment with the estimation of the creep velocity $V_L$ of the pore
vortex using Eq. (\ref{VLy}). For $V_{s}\approx 1$ cm/sec, which is typical
in the torsional oscillation experiments, we get $V_{L}\approx \allowbreak
0.7$ cm/sec. For the frequency $\nu =\allowbreak 477$ Hz in the experiment,
this corresponds to vortex displacements about $1.5\times 10^{-3}$ cm, which
is about $6$ structure periods. This looks rather satisfactory for our
scenario reducing sequences of jumps between neighboring cells to
quasi-continuous vortex motion.

\section{Conclusion}

\label{concl}

We suggested the theory of motion of 3D coreless vortices through a 2D film
covering a porous substrate. Dynamics of 3D vortices is derived from
dynamics of vortex-antivortex pairs in the 2D film. The 3D vortices move
jumping from cell to cell of the substrate structure, but it can be
described in terms of the average vortex velocity like vortex motion in
continuous media. Frequency of jumps, and vortex velocity correspondingly,
is determined by dissociation rate of vortex-antivortex pairs, which is
known from dynamics of 2D superfluid films based on the Kosterlitz-Thouless
theory. We calculated dissipation intensity and its temperature dependence
analytically and numerically. The theory is compared with the experiments on
torsional oscillations of $^{4}$He superfluid film adsorbed on a rotating
porous-glass substrate. We explain the second (rotation-induced) dissipation
peak on the temperature dependence, which was revealed in these experiments.
Quantitative comparison between theory and experiment looks satisfactory,
especially for the shape and the position of the peak.

Though we focused on application of our theory to torsional-oscillation
experiments in $^4$He films on porous-glass substrates, we believe that the
theory has a much wider area of possible applications. The concept of
quasi-continuous vortex motion in porous media with parameters determined
from dynamics of 2D superfluid films should be applicable not only to
straight vortices induced by steady rotation. For example, one may apply the
theory also to vortex rings and to the vortex tangle if they can be created
in porous media. The theory can also be used for the analysis of steady
vortex motion in various situations, e.g. in the process of heat transfer.

\section*{ACKNOWLEDGMENTS}

Authors are grateful to M. Kubota for numerous discussions and providing
necessary experimental data. S. K. N. thanks the Racah Institute of Physics
of the Hebrew University of Jerusalem for hospitality and support and
acknowledges a partial support by the grants N 05-08-01375 and 07-02-01124
of RFBR and the grant NSH-6749.2006.8 of the state support of leading
scientific schools by the President of the Russian Federation. .

\end{document}